\documentclass[
superscriptaddress,nofootinbib,amsmath,amssymb]{revtex4-2}
\usepackage[colorlinks=true,linkcolor=blue,citecolor=magenta,linktocpage=true]{hyperref}
\usepackage{graphicx}% include figure files
\usepackage{bm}% bold math
\usepackage{hyperref}% add hypertext capabilities
\usepackage{mathtools}
\usepackage{comment}

\renewcommand{\d}{{\rm d}}
\newcommand{\q}{{\qquad\qquad}}
\newcommand{\f}{\frac}

\newcommand{\mrm}[1]{_{\rm #1}}

\numberwithin{equation}{section}

\makeatletter
\def\l@subsubsection#1#2{}
\makeatother

\begin{document}

\begin{flushright}
CERN-TH-2021-105\vspace*{0.5cm}
\end{flushright}

\title{Hawking radiation by spherically-symmetric static black holes for all spins:\\
II - Numerical emission rates, analytical limits and new constraints
\vspace*{0.5cm}}

\author{Alexandre Arbey}
\email{alexandre.arbey@ens-lyon.fr}
\affiliation{Univ Lyon, Univ Claude Bernard Lyon 1,\\ CNRS/IN2P3, IP2I Lyon, UMR 5822, F-69622, Villeurbanne, France}
\affiliation{Theoretical Physics Department, CERN, CH-1211 Geneva 23, Switzerland}
\affiliation{Institut Universitaire de France (IUF), 103 boulevard Saint-Michel, 75005 Paris, France}

\author{Jérémy Auffinger}
\email{j.auffinger@ipnl.in2p3.fr}
\affiliation{Univ Lyon, Univ Claude Bernard Lyon 1,\\ CNRS/IN2P3, IP2I Lyon, UMR 5822, F-69622, Villeurbanne, France}

\author{Marc Geiller}
\email{marc.geiller@ens-lyon.fr}
\affiliation{Univ Lyon, ENS de Lyon, Univ Claude Bernard Lyon 1,\\ CNRS, Laboratoire de Physique, UMR 5672, F-69342 Lyon, France}

\author{Etera R. Livine}
\email{etera.livine@ens-lyon.fr}
\affiliation{Univ Lyon, ENS de Lyon, Univ Claude Bernard Lyon 1,\\ CNRS, Laboratoire de Physique, UMR 5672, F-69342 Lyon, France}

\author{Francesco Sartini}
\email{francesco.sartini@ens-lyon.fr}
\affiliation{Univ Lyon, ENS de Lyon, Univ Claude Bernard Lyon 1,\\ CNRS, Laboratoire de Physique, UMR 5672, F-69342 Lyon, France}

%\date{\today}

\begin{abstract}
\vspace{0.5cm}

In the companion paper \cite{Arbey:2021jif} we have derived the short-ranged potentials for the Teukolsky equations for massless spins $(0,1/2,1,2)$ in general spherically-symmetric and static metrics. Here we apply these results to numerically compute the Hawking radiation spectra of such particles emitted by black holes (BHs) in three different ansatz: charged BHs, higher-dimensional BHs, and polymerized BHs arising from models of quantum gravity. In order to ensure the robustness of our numerical procedure, we show that it agrees with newly derived analytic formulas for the cross-sections in the high and low energy limits. We show how the short-ranged potentials and precise Hawking radiation rates can be used inside the code \texttt{BlackHawk} to predict future primordial BH evaporation signals for a very wide class of BH solutions, including the promising regular BH solutions derived from loop quantum gravity. In particular, we derive the first Hawking radiation constraints on polymerized BHs from AMEGO. We prove that the mass window $10^{16}-10^{18}\,$g for all dark matter into primordial BHs can be reopened with high values of the polymerization parameter, which encodes the typical scale and strength of quantum gravity corrections.
\end{abstract}

\maketitle

\newpage

\tableofcontents

\section*{Introduction}
\label{sec:introduction}

The most striking feature of black holes (BHs) might be that, in spite of their name, they (supposedly) radiate particles and slowly evaporate, as first discovered by Hawking \cite{Hawking1975}. Hawking radiation (HR) causes BHs to lose mass, charge and angular momentum, up to a late evaporation stage where physics is so far unsettled. This late stage is sometimes conjectured to lead either to complete disappearance or to a stable remnant, but it is clear that a proper understanding of the final state of a black hole after its evaporation requires an understanding of the quantum nature of gravity.

BHs in the stellar mass range are observed \textit{e.g.}~at LIGO/VIRGO via the gravitational waves they emit when binary BHs merge \cite{LIGO2019,LIGO2020}, while shadows of supermassive BHs at the center of galaxies can be probed by large array interferometers \cite{EHT2019}. Primordial BHs (PBHs) formed in the early universe could be lighter that the Oppenheimer--Volkoff limit, down to the Planck mass. As Hawking radiation (HR) gets more energetic when the BH mass decreases, one could hope to observe these light PBHs thanks to the radiation they emit in every direction. For now, there is currently no direct observational evidence of HR, resulting in constraints on the abundance of PBHs in the universe, as those which have not totally evaporated by now ($M\mrm{BH} \gtrsim 10^{14}\,$g for a Kerr BH) account for some fraction of the dark matter (DM). We refer the reader to \cite{Carr2020} for a complete review on this topic.

Interestingly, there is still an open window in the asteroid mass range $10^{17} \lesssim M\mrm{BH} \lesssim 10^{21}\,$g for PBHs to represent all of DM, solving this long-standing issue in cosmology \cite{Katz2018,Carr2020}, although the lower part of this window may be closed by future gamma ray observatories such as AMEGO \cite{Coogan2021}. This window is precisely constrained by HR predictions. However, since different BH solutions to the Einstein equations provide different HR signals, it is worth having a new look at the constraints on the PBH abundance from various examples of spherically-symmetric static black holes.

In a companion paper \cite{Arbey:2021jif} we have derived the short-ranged potentials of the Teukolsky equations for a wide class of BH solutions, namely spherically-symmetric and static BHs, setting the mathematical background for HR computations. In the present paper, we use these potentials to compute the HR of benchmark BH solutions: charged BHs, higher-dimensional BHs and polymerized BHs. An abundant literature is dedicated to the case of charged BHs (\textit{e.g.}~\cite{Carter1974,Gibbons1975,Zaumen1974,Page3,Cvetic1998,Crispino2009,Lehman2019}) and higher-dimensional BHs (\textit{e.g.}~\cite{Kanti2002_spin0,Kanti2003_spin1_12,Harris2003,Johnson2020}, for a review see \cite{Kanti2012}), but much less work have focused on polymerized BHs \cite{Anacleto2020,Modesto2010,Alesci2011,Hossenfelder2012,Alesci2012,Bojowald2018} (see in particular \cite{Moulin2019}). We stress that our derivation of the potentials for HR applies in particular to the class of regular BHs, which are BH solutions showing no singularity at the coordinate origin. Polymerized BHs, inspired by loop quantum gravity (LQG, see \cite{Ashtekar:2004eh} for a review), are an example of regular BHs \cite{Ashtekar2006,Modesto2006,Bohmer2007,Olmedo:2017lvt,Ashtekar:2018cay,Ashetkar2018,BenAchour:2017ivq,BenAchour:2018khr,Bodendorfer:2019cyv,Bodendorfer:2019xbp,Bodendorfer:2019nvy,Bodendorfer2019,Alesci:2019pbs,Alesci:2020zfi,Sartini:2020ycs,Brahma:2020eos,Gambini:2020qhx}. Recent work on other regular BH solutions include \textit{e.g.}~\cite{Rincon2020,Berry2021,Cai2021,Baake2021,Molina2021}, some of which also discuss HR.

In this paper, we use the short-ranged potentials derived in the companion paper \cite{Arbey:2021jif} to compute numerically the HR rates for massless particles of spin 0, 1, 2 and $1/2$, thereby completing previous literature. We present in details three examples as illustrations, namely charged BHs, higher-dimensional BHs, and polymerized BHs. The important point we wish to emphasize is that most of the previous studies on HR were performed with heavy use of numerical methods due to the lack of a general derivation of the short-ranged potentials. Here we take advantage of the analytical results derived in \cite{Arbey:2021jif}. As a consistency check, we also compare our results at low and high energy with analytical limits, some of which are derived for the first time here. This gives strong support for the robustness of our numerical method based on short-ranged potentials, which can then be used to predict HR radiation signals for other BH metrics for which there is no analytical results yet. During this study, we have also modified the public code \texttt{BlackHawk} \cite{blackhawk}, written by a subset of the present authors, to produce the Hawking spectra of BHs with the new metrics. This improvement will be part of a forthcoming update of \texttt{BlackHawk}. As a novel application, we use the computed emission rates for polymerized BHs to reassess the MeV-GeV photon constraint on PBHs from AMEGO, and show that a sufficiently high value of the polymerization parameter $\varepsilon$ reopens the mass window $M\mrm{PBH} \lesssim 10^{18}\,$g for all DM in the form of PBHs. This is the first constraint ever set on polymerized BHs with Hawking radiation.

The paper is organized as follows. Section~\ref{sec:starting_eqs} briefly reviews the results of the companion paper and gives the important analytical formulas necessary to the study of Hawking radiation. In Section~\ref{sec:metrics}, we introduce the BH metrics on which we focus, motivate their choice and physical relevance. Section~\ref{sec:HR} presents our main results, with the Hawking radiation numerical computations and the new constraint on polymerized primordial black holes. Finally we conclude and give perspectives for future work.

%%%%%%%%%%%%
\section{General equations}
\label{sec:starting_eqs}
%%%%%%%%%%%%

In this Section we summarize the results of the companion paper \cite{Arbey:2021jif}, which establishes the mathematical framework necessary for the the study of Hawking radiation emitted by spherically-symmetric static black holes. We introduce the general metric ansatz and recall the main analytical formulas: short-ranged potentials, Hawking radiation cross-section with low and high energy limits.

%%%%
\subsection{Short-ranged potentials}
%%%%

We are interested in spherically-symmetric and static metrics, which in Boyer--Lindquist coordinates take the general form
\begin{equation}
    \d s^2 = -G(r)\d t^2 + \dfrac{1}{F(r)}\d r^2 + H(r)\d \Omega^2\,,\label{eq:metric}
\end{equation}
where $\d\Omega^2 = \d\theta^2 + \sin(\theta)\d\varphi^2$.\footnote{From now on, we use natural units such that $G = c = \hbar = k\mrm{B} = 4\pi\varepsilon_0 = 1$.} We restrict ourselves to black hole solutions, by which here we mean metrics that are asymptotically flat
\begin{equation}\label{falloffs}
	F(r)\underset{r\rightarrow+\infty}{\longrightarrow}1\,,\q G(r)\underset{r\rightarrow+\infty}{\longrightarrow}1\,,\q H(r)\underset{r\rightarrow+\infty}{\sim}r^2\,,
\end{equation}
and present a horizon at some radius $r\mrm{H}$ which is a pole of $F$. For such metrics the ADM mass of the BH is then \cite{Gourgoulhon2007}
\begin{equation}
    M = \underset{r\rightarrow+\infty}{\lim} \,\,\dfrac{1}{2}\left( \frac{r}{F} + \dfrac{H}{r} - \partial_r H\right).
\end{equation}
We have shown in the companion paper \cite{Arbey:2021jif} that for these metrics the equations of motion of spin 0, 1, 2 and $1/2$ massless particles can be transformed into one-dimensional Schr\"odinger-like wave equations
\begin{equation}
    \partial_*^2 Z + \Big(\omega^2 - V\big(r(r^*)\big)\Big)Z = 0\,,\label{eq:schrodinger}
\end{equation}
where the tortoise coordinate is defined as
\begin{equation}
    \dfrac{\d r^*}{\d r} \equiv \dfrac{1}{\sqrt{FG}}\,,\label{eq:tortoise}
\end{equation}
and we have introduced the notation $\partial_* \equiv \partial_{r^*}$. These tortoise coordinates $r^*(r)$ are explicitly computed for the BHs of interest in Appendix \ref{sec:tortoise}. The potentials computed in \cite{Arbey:2021jif} reduce to
\begin{subequations}\label{eq:potentials}
\begin{align}
	&V_0(r^*)=\nu_0\f{G}{H}+\f{\partial_*^2\sqrt{H}}{\sqrt{H}}\,,\\
	&V_1(r^*)=\nu_1\f{G}{H}\,,\\
	&V_2(r^*)=\nu_2\f{G}{H}+\f{(\partial_*H)^2}{2H^2}-\f{\partial_*^2\sqrt{H}}{\sqrt{H}}\,,\\
	&V_{1/2}(r^*)=\nu_{1/2}\f{G}{H} \pm \sqrt{\nu_{1/2}}\,\partial_*\left( \sqrt{\f{G}{H}} \right),
\end{align}
\end{subequations}
where the intermediate $r$ dependency is not shown for conciseness. The spin-dependent parameters $\nu_i$ are given by $\nu_0 = l(l+1) = \nu_1$, $\nu_2 = l(l+1)-2$ and $\nu_{1/2} = l(l+1)+1/4$ where $l = s, s+1,\dots$ is the total angular momentum of the wave and $m = -l,\dots,+l$ its projection. The shape of the potentials and their discrepancies with the reference Schwarzschild case as functions of the parameters of the various metric models are detailed in the companion paper \cite{Arbey:2021jif}.

\subsection{Greybody factors, low and high energy limits}

With the short-ranged potentials at our disposal, it is numerically straightforward to compute the greybody factors needed to estimate the Hawking radiation rate. Hawking radiation \cite{Hawking1975} is a semi-classical phenomenon of quasi-thermal emission of particles by BH horizons. The rate of emission of one degree of freedom $i$ per unit time $t$ and energy $E$ is given by
\begin{equation}
    \dfrac{\d^2 N_i}{\d t\,\d E} = \sum_{l,m}\dfrac{1}{2\pi}\dfrac{\Gamma_i(E,M,x_j)}{e^{E/T}-(-1)^{2s_i}}\,,\label{eq:master}
\end{equation}
where $s_i$ is the spin of the particle $i$ and $T$ is its Hawking temperature given by
\begin{equation}
    T(M,x_j) = \dfrac{\kappa}{2\pi}\,,
\end{equation}
where $\kappa$ is the surface gravity of the BH. This latter is obtained from the formula
\begin{equation}
    \kappa^2 \equiv \left.-\dfrac{1}{2}\nabla_\mu k_\nu \nabla^\nu k^\mu\right|_{\rm hor} = \left.\dfrac{1}{4}\frac{FG^{\prime 2}}{G}\right|_{\rm hor}\,,
\end{equation}
where $k^\mu = (1,0,0,0)$ is the timelike Killing vector and ``hor'' denotes the horizon $r=r_\text{H}$. The greybody factor is the probability that a particle generated by thermal fluctuations at the horizon escapes to spatial infinity. If we consider a wave function which is purely ingoing on the horizon, \textit{i.e.}~with
\begin{equation}
    Z(r^*) \underset{r^*\rightarrow -\infty}{\sim} A\mrm{hor}^{\rm in} e^{-i\omega r^*}\,,
\end{equation}
and given at infinity by
\begin{equation}
    Z(r^*) \underset{r^*\rightarrow +\infty}{\sim} A\mrm{\infty}^{\rm in}e^{-i\omega r^*} + A\mrm{\infty}^{\rm out}e^{+i\omega r^*}\,,
\end{equation}
then the greybody factor is obtained as
\begin{equation}
    \Gamma_i(E,M,x_j) = \left|\dfrac{A\mrm{hor}^{\rm in}}{A\mrm{\infty}^{\rm in}}\right|^2\,.
\end{equation}
In general, the greybody factor and the temperature (or the surface gravity) depend on the BH mass $M$ but also on the precise shape and parameters of the metric, that is the set of $x_j$. As we have decomposed the wave in spin-weighted spherical harmonics to obtain the radial potentials in \cite{Arbey:2021jif}, the greybody factors $\Gamma_i$ also depend on the particle $i$ angular momentum parameters $(l,m)$. The spherical symmetry reduces the sum on $m$ in Eq.~\eqref{eq:master} to a factor $2l+1$. To compute these greybody factors, we have used the same kind of \texttt{Mathematica} scripts that are given with the public code \texttt{BlackHawk} \cite{blackhawk}. Finally, we also define the cross-section $\sigma_i$ for a particle $i$ by
\begin{equation}
    \sigma_i \equiv \dfrac{\pi}{E^2}\Gamma_i\,.
\end{equation}
We emphasize here that the new HR results obtained in this paper will be part of a forthcoming update of \texttt{BlackHawk}, where the aforementioned scripts and tabulated greybody factors will be publicly available.

\subsubsection{High energy limit}

The high energy limit is usually called the ``geometrical optics'' approximation, because fields of all spins experience the BH as an optical obstacle whose extension is given by the effective area $A\mrm{eff}(x_j)$ enclosed by the last unstable circular orbit. This area depends on the set of parameters $x_j$ of the BH metric. Let $b\mrm{c}$ be the critical impact parameter for which the incoming massless fields would reach an unstable circular orbit of radius $r\mrm{c}$. For a general BH metric of the form \eqref{eq:metric}, this is called the ``photon sphere", defined as the innermost unstable circular orbit for a massless test particle in rotation around the black hole. Photons follow null geodesics, meaning that for an affine parameter $\lambda$ we have
\begin{equation}
    0= g_{\mu\nu}\f{\d x^\mu}{\d \lambda}\f{\d x^\nu}{\d \lambda}\,. \label{eq:null_geodesic}
\end{equation}
Along every geodesic there are two conserved quantities, the energy $E$ and the angular momentum $L$, associated respectively to the Killing vector fields $\partial_t $ and $\partial_\varphi$. Using the metric ansatz~\eqref{eq:metric} they are given by
\begin{equation}
    E \equiv G \f{\d t}{\d \lambda}\,,\qquad\qquad
    L \equiv H \f{\d \varphi}{\d \lambda}\,.
    \label{eq:geodesics_killing}
\end{equation}
Inserting these expressions into Eq.~\eqref{eq:null_geodesic} and choosing a planar orbit at $\theta=\pi/2$, we get
\begin{equation}
    \left (\f{\d r}{\d \lambda}\right)^2 =\frac{1}{F} \left( \frac{E^2}{G}-\f{L^2}{H}\right) \equiv -V_{\rm{eff}} \,.
\label{eq:geodesic_effective}
\end{equation}
The radial acceleration is then given by $\dfrac{\d^2 r}{\d \lambda^2} = - V_{\rm{eff}}^\prime$. Let us remark that we could obtain the same result by directly calculating the radial geodesic equation. In order to have an unstable circular orbit there must be a critical radius $r\mrm{c}$ such that $V_{\rm{eff}}^\prime(r\mrm{c})=0$ and $V_{\rm{eff}}^{\prime\prime}(r\mrm{c})<0$. Since on this orbit the radial velocity must be vanishing, we can also use Eq.~\eqref{eq:geodesic_effective} to constrain the energy and the angular momentum to satisfy
\begin{equation}
   V_{\rm{eff}}(r\mrm{c})=0\,\Rightarrow \f{L^2}{H(r\mrm{c})} = \f{E^2}{G(r\mrm{c})}\,.
\end{equation}
It is then straightforward to verify that the condition for the unstable orbit reduces to
\begin{subequations}\label{eq:conditions}
\begin{align}
    V_{\rm{eff}}^\prime(r\mrm{c})=0\quad &\Rightarrow\quad\dfrac{G^\prime(r\mrm{c})}{H^\prime(r\mrm{c})} - \dfrac{G(r\mrm{c})}{H(r\mrm{c})} = 0\,,\\
     V_{\rm{eff}}^{\prime\prime}(r\mrm{c})<0\quad &\Rightarrow\quad H(r\mrm{c}) G^{\prime\prime}(r\mrm{c})- H^{\prime\prime}(r\mrm{c}) G(r\mrm{c}) < 0\,.
\end{align}
\end{subequations}
The critical impact parameter for a massless particle is then defined with respect to the energy and the angular momentum as
\begin{equation}
    b_{\rm{c}}^2 \equiv \f{L^2}{E^2}=\f{H(r\mrm{c})}{G(r\mrm{c})}\,.
\end{equation}
Finally, the effective area (classical scattering) of the BH is given in $4+n$ dimensions as
\begin{equation}
    \sigma_\infty = A\mrm{eff}(x_j) = \dfrac{\pi^{(n+2)/2}b\mrm{c}^{n+2}}{\Gamma\big((n+4)/2\big)}\,,\label{eq:sigma_infty_tr}
\end{equation}
where $\Gamma$ is the Euler gamma function.

For the particular example of a so-called $tr$-symmetric metric, for which we have $F(r) = G(r) \equiv h(r)$ and $H(r) = r^2$ in Eq.~\eqref{eq:metric}, the conditions~\eqref{eq:conditions} reduce to \cite{Decanini2010,Decanini2011}
\begin{equation}
    h^\prime(r\mrm{c}) - \dfrac{2}{r\mrm{c}}h(r\mrm{c}) = 0\,,\qquad h^{\prime\prime}(r\mrm{c}) - \dfrac{2}{r\mrm{c}^2}h(r\mrm{c}) < 0\,.\label{eq:trsymmetric_condition}
\end{equation}
Then, the impact parameter is given by
\begin{equation}
    b\mrm{c} = \dfrac{r\mrm{c}}{\sqrt{h(r\mrm{c})}}\,.
\end{equation}
In the Schwarzschild case, we obtain for all particle spins (see \textit{e.g.}~\cite{MacGibbon1990} and references therein)
\begin{equation}
    \sigma_\infty = 27\pi M^2 \equiv \sigma\mrm{GO}\,,\label{eq:sigma_GO_S}
\end{equation}
which is the usual geometric approximation cross-section.

\subsubsection{Low energy limit}

At low energy, massless fields of each spin behave differently. Several methods have been used to obtain the low energy limits for the cross-section. Classical scattering arguments apply to the low energy limit of the spin 0 field, as can be found in \cite{Anacleto2020} which uses the partial wave decomposition and the small angle approximation.

More generally, authors use a ``matching'' method which consists in reducing the radial spin-dependent Teukolsky equation (see \cite{Arbey:2021jif}) to a simplified version in the ``far field" region ($r\rightarrow+\infty$) and in the ``horizon" region ($r\rightarrow r\mrm{H}$). Depending on the number of poles (number of horizons) of the metric components $F$ and $G$, the differential equation obtained is some form of a Heun equation. This dependency on the spin makes it difficult to provide a general procedure to obtain the desired equation, especially in the general case $F\ne G$ studied here. Then, solving in each region independently, while applying the correct boundary conditions, and matching the two solutions in the intermediate region gives an analytical expression for the greybody factor. Arguments that justify the matching procedure are given \textit{e.g.}~in \cite{Cvetic1998}. This expression can be expanded in the limit $\omega r\mrm{H} \rightarrow 0$ to obtain the analytical low energy limit for the cross-section. One further simplification is that at low energy, only the lowest momentum partial wave $l = s$ participates significantly to the result. This is the method used \textit{e.g.}~in \cite{Cvetic1998,Kanti2002_spin0,Kanti2003_spin1_12,Harris2003,Motohashi2021}.
In the Schwarzschild case (denoted by a superscript S), the low energy limits for the various spins studied here are (see \textit{e.g.}~\cite{MacGibbon1990} and references therein)
\begin{equation}\label{eq:low_energy_S}
\begin{array}{ll}
    \sigma_{0}^{\rm S} = 4r\mrm{S}^2\,,&\q
    \sigma_{1}^{\rm S} = \dfrac{4}{3}r\mrm{S}^4E^2\,,\\
    \sigma_{2}^{\rm S} = \dfrac{4}{45}r\mrm{S}^6E^4\,,&\q
    \sigma_{1/2}^{\rm S} = \dfrac{1}{2}r\mrm{S}^2\,.
\end{array}
\end{equation}

\section{Examples of metrics}
\label{sec:metrics}

To illustrate our results, we have selected three different metrics for which we will compute the HR : $i)$ charged BHs, $ii)$ higher-dimensional BHs, $iii)$ polymerized BHs arising from loop quantum gravity. In this Section we briefly present the corresponding metrics and the parameters $x_j$ which they contain.

\subsection{Charged black holes}

The no-hair theorem states that a BH (in general relativity coupled to electromagnetism) is entirely characterized by its mass $M$, angular momentum $J$ and electric charge $Q$. Astrophysical BHs are expected to have a sizeable spin due either to their formation mechanism through the collapse of stars (for stellar BHs) or due to long-term accretion of orbiting gas clouds (for supermassive BHs at the center of galaxies). These are however not expected to have a sizeable electric charge because their environment of formation is typically electrically neutral. However, primordial BHs can have either zero spin (if formed during radiation domination era) or high spin (if formed during matter domination era) \cite{Garcia2020,Burke2020,DeLuca2020,Arbey2020_spin}. The spin is bounded to $a \equiv J/M < M$ to avoid the breaking of the horizon and the appearance of a naked singularity. HR makes this spin decrease in time because the emission of particles with an angular momentum aligned to the BH spin is preferred. The paradigm is exactly the same for electric charge. The electric charge is bounded to\footnote{The general relation for a Kerr--Newman (charged and rotating BH) is $a^2 + Q^2 < M^2$.} $Q < M$ and decreases in time because emission of particles of the same charge as the BH is preferred \cite{Carter1974,Gibbons1975,Zaumen1974,Page3}. The universe is supposed to be neutral at all times, but processes in the early universe could produce electrically charged regions that collapse into electrically charged PBHs (astrophysical BHs are generally assumed to be neutral). The charge can be maintained until the present epoch if it was initially very close to the extremal case $Q \lesssim M$, with the same kind of evolution as described for the spin in \cite{Arbey2020_spin}. Random fluctuations of the charge would nevertheless remain until decay has reached the Planck mass \cite{Page3,Lehman2019}.

Our formalism is not adapted to treat the emission of particles that have (electromagnetic) couplings with the background metric; therefore we did not compute the Hawking rates of charged particles in this particular example. With our notation for the general family \eqref{eq:metric}, the metric of a Reissner--Nordstr\"om BH is \cite{Arbey:2021jif}
\begin{equation}
    \begin{matrix}
		F(r) = G(r) = 1 - \dfrac{r\mrm{S}}{r} + \dfrac{r_Q^2}{r^2} = \dfrac{(r-r_+)(r-r_-)}{r^2} \,,\qquad H(r) = r^2\,,\label{eq:metric_charged}
	\end{matrix}
\end{equation}
where $r\mrm{S} \equiv 2M$ is the Schwarzschild radius and $r_Q^2 = Q^2$ in our system of units. The temperature is given by
\begin{equation}
    T_Q(M,Q) = \dfrac{\kappa_Q}{2\pi} = \dfrac{r_+ - r_-}{4\pi r_+^2}\,,\label{eq:temp_charged}
\end{equation}
where the horizon radii are
\begin{equation}
    r_\pm(M,Q) \equiv M\left( 1 \pm \sqrt{1 - \dfrac{Q^2}{M^2}} \right).\label{eq:rpm_charged}
\end{equation}
In the limit $Q\rightarrow 0$ we recover the Schwarzschild case with $r_+ = r\mrm{S}$ and $r_- = 0$. In the opposite limit $Q\rightarrow M$ we obtain $r_+ = r_-$ and $T_Q = 0$, which means that there is no Hawking emission; the so-called \textit{extremal} BH is eternal. Overall, the temperature---and thus the emission power---decreases as $Q$ increases, because $r_- \rightarrow r_+$. The result of integration of Eq.~\eqref{eq:tortoise} for a charged BH is given in Appendix~\ref{sec:tortoise}.

%%%%%
\subsection{Higher-dimensional black holes}
%%%%%

Although general relativity is typically studied in four spacetime dimensions, which coincides fairly well with astronomical observations, small extra spatial dimensions are not ruled out by particle physics experiments, nor by the propagation of gravitational waves (see \cite{Johnson2020} and references therein). As in \cite{Johnson2020}, we will consider here \textit{large} extra dimensions, that is to say dimensions with typical size larger than the Planck size $R \gg \ell\mrm{P}$, and \textit{small} BHs, that is to say BHs with horizon radius $r\mrm{H} \ll R$. Hawking radiation on the bulk and in the brane of such BHs was previously studied in \textit{e.g.}~\cite{Harris2003}. On the 4-dimensional brane, the metric is
\begin{equation}
        F(r) = G(r) = h(r) \equiv 1 - \left(\dfrac{r\mrm{H}}{r}\right)^{n+1}, \quad H(r) = r^2\,,\label{eq:metric_higher}
\end{equation}
where $n>0$ is the number of extra dimensions and the horizon radius is given by
\begin{equation}
    r\mrm{H} = \dfrac{1}{\sqrt{\pi}M_*}\left( \dfrac{M}{M_*} \right)^{1/(n+1)}\left( \dfrac{8\Gamma\big((n+3)/2\big)}{n+2} \right)^{1/(n+1)},
\end{equation}
where $\Gamma$ is the Euler gamma function and the rescaled Planck mass is $M\mrm{Pl}^2 = M_*^{n+2} R^{n}$. The temperature is
\begin{equation}
    T_n = \dfrac{\kappa_n}{2\pi} = \dfrac{n+1}{4\pi r\mrm{H}}\,.\label{eq:temp_higher}
\end{equation}
We recover the Schwarzschild results for $n = 0$. The result of integration of Eq.~\eqref{eq:tortoise} for a higher-dimensional BH is given in Appendix~\ref{sec:tortoise}.

%%%%%
\subsection{Polymerized black holes}
%%%%%

Polymerized BHs have emerged as an effective template for black holes in loop quantum gravity\footnote{The term ``polymerized" refers to the polymer-like quantization scheme inherited from loop quantum gravity \cite{Ashtekar:2004eh}.}. They are studied by applying the techniques of loop quantum gravity and loop quantum cosmology \cite{Ashtekar:2011ni} to mini-superspace black hole spacetimes. Although many models have been proposed in the literature (see \textit{e.g.}~the non-exhaustive list \cite{Ashtekar2006,Modesto2006,Bohmer2007,Olmedo:2017lvt,Ashtekar:2018cay,Ashetkar2018,BenAchour:2017ivq,BenAchour:2018khr,Bodendorfer:2019cyv,Bodendorfer:2019xbp,Bodendorfer:2019nvy,Bodendorfer2019,Alesci:2019pbs,Alesci:2020zfi,Sartini:2020ycs,Brahma:2020eos,Gambini:2020qhx}), depending on the details of the regularization of the Hamiltonian, here we focus for definiteness on the particular class of effective metrics derived in \cite{Modesto:2008im,Modesto:2009ve}. These metrics are regular, \textit{i.e.}~do not admit a singularity at $r=0$, and remain asymptotically flat. The resolution of the singularity arises from effects of quantum geometry which become relevant at the Planck scale. Here we treat these metrics as phenomenological ansatz for BHs taking into account loop quantum gravity corrections at the semi-classical level.

Identifying possible signatures of quantum gravity has recently regained interest due to the increasing precision in the detection of gravitational waves. Indeed, the last phase of BH merging, known as the ringdown phase, is extremely sensitive to the details of the metric structure of BHs. The deformed BH resulting from the coalescence of two BHs settles down to a stable (axially) symmetric shape by emitting gravitational waves of defined frequencies known as quasi-normal modes. The determination of these modes for a given metric and the comparison with the ringdown signal could discriminate between different models of BHs \cite{Barrau2019}.

The stability of polymerized BHs, the equations governing their geometry, and their Hawking radiation were studied \textit{e.g.}~in \cite{Modesto2010,Alesci2011,Hossenfelder2012,Alesci2012,Bojowald2018,Moulin2019,Anacleto2020} for scalar waves and gravitational waves. The results for spin 1 are new to this paper. In reference \cite{Moulin2019} one can find some results for the massless field of spin $1/2$. Our results for this case however differ quantitatively from \cite{Moulin2019}, as we will see below. It is also noteworthy that no constraint has yet been derived for these polymerized BHs using Hawking radiation. For the first time, we give such a constraint on the primordial black holes abundance as a dark matter component in Section~\ref{sec:constraint}.

We now turn to the precise description of the model. The polymerized BH metric studied in \cite{Modesto:2008im,Modesto:2009ve} has defining functions given by
\begin{subequations}\label{eq:metric_LQG}
\begin{align}
    &G= \dfrac{(r - r_+)(r - r_-)(r + \sqrt{r_+r_-})^2}{r^4 + a_0^2}\,,\\
    &F= \dfrac{(r - r_+)(r - r_-)r^4}{(r + \sqrt{r_+r_-})^2(r^4 + a_0^2)}\,,\\
    &H= r^2 + \dfrac{a_0^2}{r^2}\,.
\end{align}
\end{subequations}
Here $a_0$ and $\varepsilon$ are the two parameters encoding the quantum gravity deformation from the Schwarzschild metric. The parameter $a_0$ is the minimal area in loop quantum gravity, also referred to as the {\it area gap}. It is typically of the Planck scale. The deformation parameter $\varepsilon \ge 0$ is an \textit{a priori} independent parameter indicating the typical scale of the geometry fluctuations in the Hamiltonian constraints of the theory as they get renormalized from the Planck scale to astrophysical scales. Although one can be tempted to keep it very small $\varepsilon \ll 1$, nothing \textit{a priori} forbids it from growing large and it can be interesting to consider possible high values of this deformation parameter in order to understand the effects of loop quantum gravity corrections.

The two roots of the metric components are $r_+ = 2m$ and $r_- = 2mP(\varepsilon)^2$. This identifies the first important effect of the polymerization: the black hole, even without electric charge, acquires a Cauchy horizon at $r=r_-$ on top of the event horizon at $r=r_+$. If the deformation parameter $\varepsilon$ is sent to 0, the radius $r_-$ is also sent to 0 (even if the area gap remains non-vanishing) and we recover a Schwarzschild-like metric. On the other hand, as $\varepsilon$ grows large, $r_-$ grows to $r_+$ (although always remaining smaller) and the polymerized BH geometrically behaves as if it carried a non-vanishing charged energy-momentum tensor although it of course does not create an electromagnetic field.
Finally, the mass parameter\footnote{We stress that the effective mass quantity $m$ defined here has nothing to do with the angular momentum projection $m$ used earlier; as we study spherically-symmetric BHs, the angular momentum projection plays no particular role.} $m$ is related to the ADM mass $M$ by $M = m(1+P)^2$ where the polymerisation factor is
\begin{equation}
    P = \dfrac{\sqrt{1 + \varepsilon^2} - 1}{\sqrt{1 + \varepsilon^2} + 1}\,.
\end{equation}
The temperature is
\begin{equation}
    T\mrm{LQG} = \dfrac{\kappa\mrm{LQG}}{2\pi} = \dfrac{4m^3(1-P^2)}{32\pi m^4+ 2\pi a_0^2} = \dfrac{r_+^2(r_+ - r_-)}{4\pi(r_+^4 + a_0^2)}\,.
\end{equation}
We can see that when $\varepsilon \ll 1$, the change in the temperature is quite negligible compared to the Schwarzschild case. We thus expect rates of emission for polymerized BHs close to the classical case. However, we see that in the limit $\varepsilon \rightarrow +\infty$, the radii collapse $r_- \rightarrow r_+$ and the temperature goes to $T\mrm{LQG} \rightarrow 0$, cancelling Hawking radiation, similarly to the charged BH with $Q\rightarrow M$. We thus expect a close behaviour of the emission rates: overall, the temperature---and thus the emission power---would decrease when $\varepsilon$ increases. What we also remark is that unless the BH has a horizon radius $r_+^2 \gtrsim a_0 $, the value of $a_0$ would have a marginal effect on the temperature (and emission rates). The result of integration of Eq.~\eqref{eq:tortoise} for a polymerized BH is given in Appendix~\ref{sec:tortoise}.

\section{Results and discussion}
\label{sec:HR}

In this Section, we show the Hawking radiation of BHs with the different metrics of Section~\ref{sec:metrics}. This is the first time Hawking radiation signals from these metrics are compared, in particular in the spin 1 massless (photon) and spin $1/2$ massless (Weyl neutrino) case of polymerized BHs emission. As literature already exists for the case of charged Reissner--Nordstr\"om BHs \cite{Page3} and higher-dimensional Schwarzschild BHs \cite{Kanti2002_spin0,Kanti2003_spin1_12,Harris2003,Johnson2020}, we put the full spectra of Hawking radiation for those in the Appendix~\ref{app:results} (see Figs.~\ref{fig:HR_charged} and \ref{fig:HR_higher}).

We confront our results to existing literature on analytical cross-sections when available to conclude on the validity of our potentials and the precision of the numerical computation. To compare our results to the limit at high energies, we define the quantity
\begin{equation}
    \beta_\infty \equiv \dfrac{\sigma_\infty}{\sigma\mrm{GO}} = \dfrac{A\mrm{eff}(x_j)}{(27/4)\pi r\mrm{S}^2}\,,
\end{equation}
where $\sigma\mrm{GO}$ is given in Eq.~\eqref{eq:sigma_GO_S}. In the low energy limit, the asymptotic expression for the cross-section depends on the spin $s$ of the radiated field. We thus define the quantities
\begin{equation}
    \beta_s \equiv \dfrac{\sigma_{s,{\rm low}}}{\sigma_{s,{\rm S}}}\,,
\end{equation}
where the low energy limits in the Schwarzschild case are given in Eqs.~\eqref{eq:low_energy_S}. We fit our numerical data at low and high energies to obtain the coefficients $\beta_\infty$ and $\beta_s$. We also check that the exponent of the energy dependency corresponds to the expected one up to $0.01\%$ precision (0 for spins 0 and $1/2$, 2 for spin 1 and 4 for spin 2). The fitting of the high energy constant limit is complicated by the oscillatory behaviour of the cross-section at high energies (see full spectra in the Appendix~\ref{app:results}). As we reach the asymptotic value from below for spin 2, the fitting procedure always slightly underestimates the value of $\beta_\infty$, but the coherence with other spins results is clear. In all 3 examples presented here, we verify that when the extra-Schwarzschild parameters go to 0 (charge, number of extra dimensions, polymerization parameter) we recover the Schwarzschild results, that is $\beta_\infty = 1$ and $\beta_s = 1$.

\subsection{Charged black holes}

The full Hawking radiation spectra of charged BHs described by the metric \eqref{eq:metric_charged} for massless fields with spins 0, 1, 2 and $1/2$ is shown in Fig.~\ref{fig:HR_charged} of Appendix~\ref{app:results}. We checked that our results are consistent with existing literature \cite{Page3}. As expected, an increase in the charge $Q$ leads to a decrease in the temperature and thus in the emitted power: the higher the BH charge, the smaller the HR rate; with the emission peaking at a smaller energy.

\begin{figure}[h]
    \centering
    \includegraphics{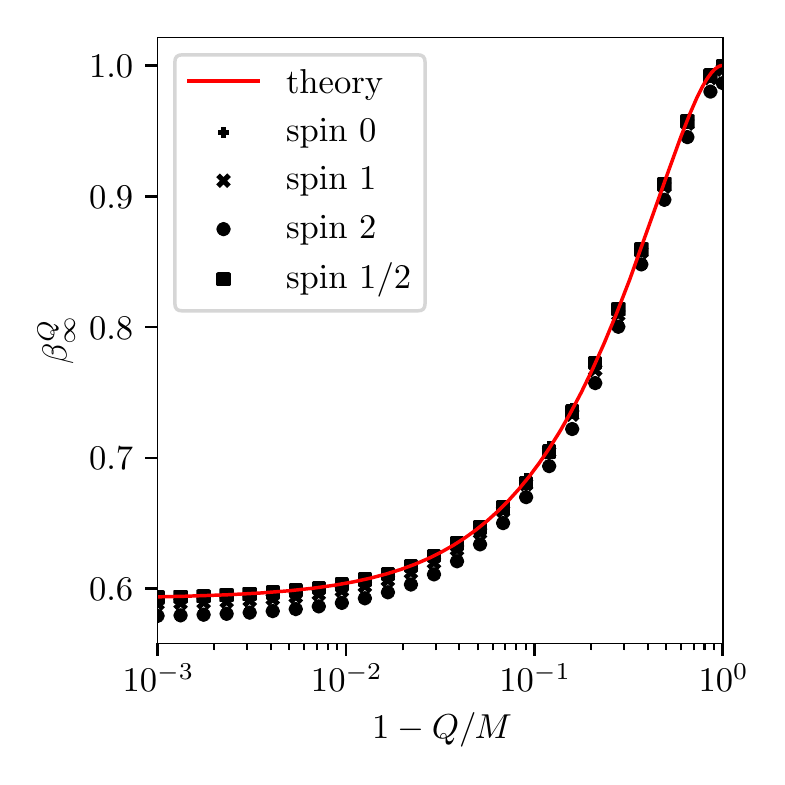}
    \includegraphics{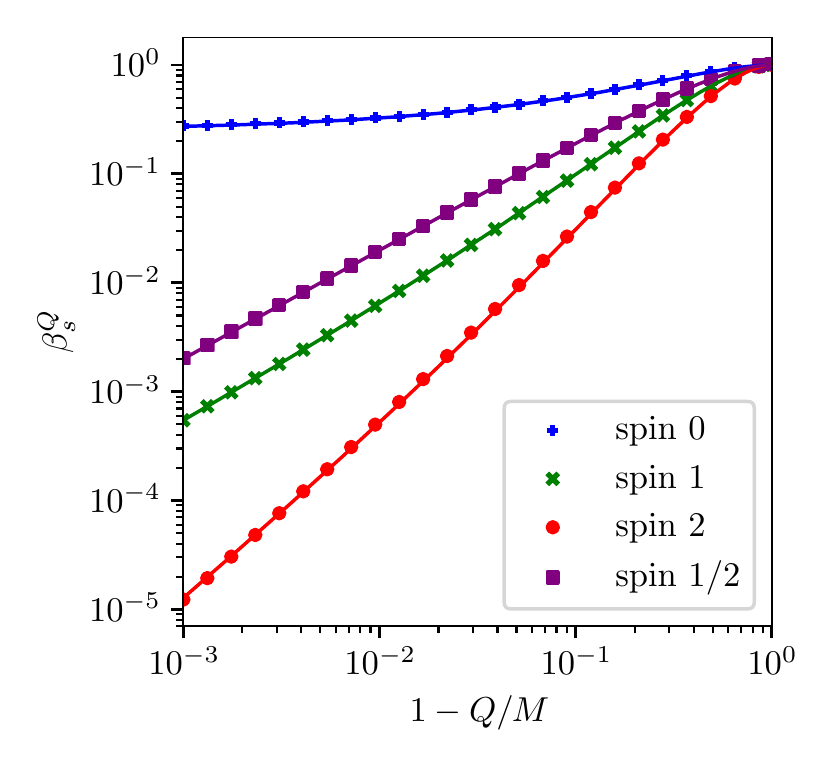}\vspace{-0.5cm}
    \caption{\textbf{Left:} Comparison between the high energy asymptotic cross-section of Eq.~\eqref{eq:charged_high_energy} for charged BHs (red solid line) and our computed results for all spins (black markers). \textbf{Right:} Comparison between the low energy asymptotic cross-section of Eqs.~\eqref{eq:charged_low_energy} (solid lines) and our computed results (markers) for all spins. Be careful of the inverted log-scale $x$-axis.}
    \label{fig:charged_asymptotics}
\end{figure}

At high energy and for a Reissner--Nordstr\"om BH with charge $Q$, it was shown in \textit{e.g.}~\cite{Crispino2009} that the asymptotic limit for the cross-section for all spins is
\begin{equation}
    \beta_\infty^Q = \dfrac{(3M+\sqrt{9M^2-8Q^2})^4}{54r\mrm{S}^2(3M^2-2Q^2+M\sqrt{9M^2-8Q^2})}\,,\label{eq:charged_high_energy}
\end{equation}
which can also be obtained from Eq.~\eqref{eq:sigma_infty_tr}. This quantity, compared to our results, is shown in Fig.~\ref{fig:charged_asymptotics} (left panel). At low energies, the general study of \cite{Cvetic1998} applies and we can predict that the low frequency limits are for all spins
\begin{subequations}
\begin{align}
    &\beta_0^Q = r_+^2/r\mrm{S}^2\,, \\
    &\beta_1^Q = r_+^2(r_+ - r_-)^2/r\mrm{S}^4\,, \\
    &\beta_2^Q = r_+^3(r_+ - r_-)^3/r\mrm{S}^6\,, \\
    &\beta_{1/2}^Q = (r_+ - r_-)^2/r\mrm{S}^2 \,.
\end{align}\label{eq:charged_low_energy}
\end{subequations}
The comparison to our results is shown in Fig.~\ref{fig:charged_asymptotics} (right panel). From these comparisons with theoretical limits at high and low energy, we can conclude that our numerical computation of the HR from charged BHs is very satisfying.

\subsection{Higher-dimensional black holes}

The full HR spectra for massless fields of spins 0, 1, 2 and $1/2$ in the case of higher-dimensional Schwarzschild BHs described by the metric \eqref{eq:metric_higher} are given in the Appendix~\ref{app:results} (see Fig.~\ref{fig:HR_higher}). The general trend is an increase in the horizon temperature with an increasing number of large extra dimensions $n$ (see Eq.~\eqref{eq:temp_higher}), resulting in more energetic HR. One unusual feature is that the oscillatory behaviour of the cross-section at high energies is damped. We have checked for consistency with the spectra of \cite{Harris2003,Johnson2020}.

\begin{figure}[h]
    \centering
    \includegraphics{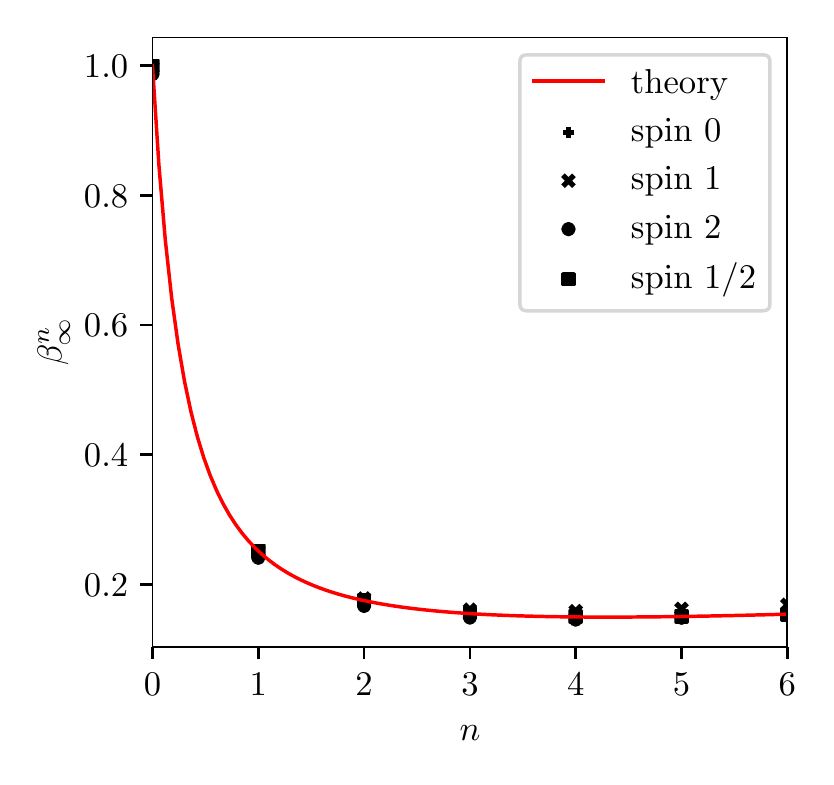}
    \includegraphics{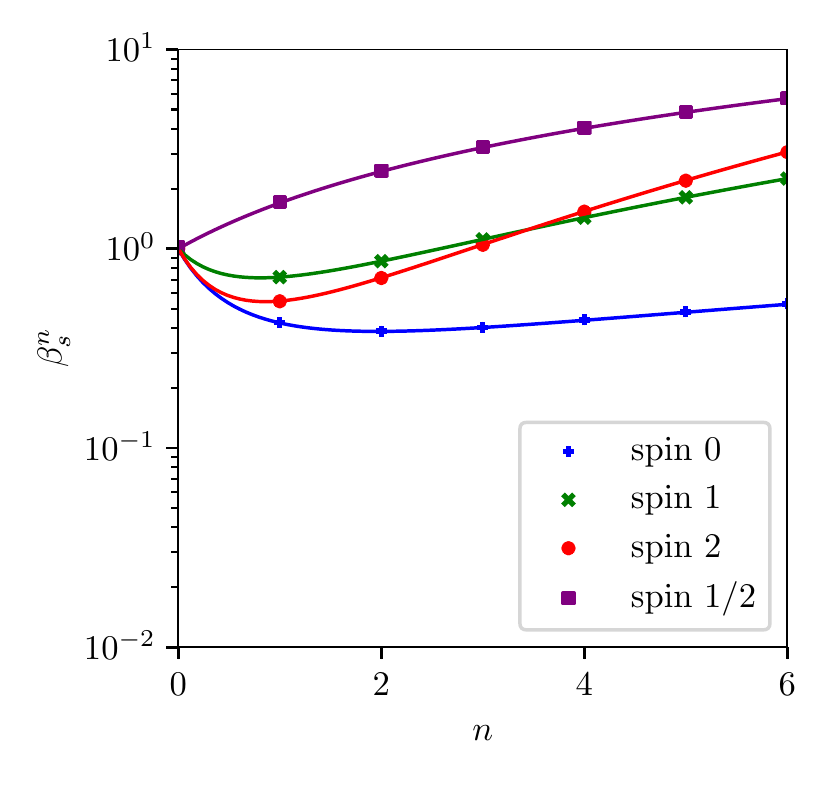}\vspace{-0.5cm}
    \caption{\textbf{Left:} Comparison between the high energy asymptotic cross-section of Eq.~\eqref{eq:higher_high_energy} for higher-dimensional BHs (red solid line) and our computed results for all spins (black markers). \textbf{Right:} Comparison between the low energy asymptotic cross-section of Eqs.~\eqref{eq:higher_low_energy} for higher-dimensional BHs (solid lines) and our computed results for all spins (markers).}
    \label{fig:higher_asymptotics}
\end{figure}

In \cite{Harris2003} (and references therein) it is shown that for every dimension $n$, the limiting value of the effective area of the horizon is $A_{\rm eff}(n) = 4\pi r\mrm{c}^2$ where
\begin{equation}
    r\mrm{c} \equiv \left(\dfrac{n+3}{2}\right)^{1/(n+1)}\sqrt{\dfrac{n+3}{n+1}}r\mrm{H}\,,
\end{equation}
which implies that the high energy cross-section satisfies 
\begin{equation}
    \beta_\infty^n = \left( \dfrac{r\mrm{H}}{r\mrm{S}} \right)^2\left( \dfrac{n+3}{2} \right)^{2/(n+1)}\left( \dfrac{n+3}{n+1} \right), \label{eq:higher_high_energy}
\end{equation}
which could also have been obtained thanks to Eq.~\eqref{eq:sigma_infty_tr}. The agreement between the theory and our results is shown in Fig.~\ref{fig:higher_asymptotics} (left panel). On the other hand, at low energy we obtain cross-sections compatible with \cite{Kanti2002_spin0,Kanti2003_spin1_12,Harris2003}
\begin{subequations}\label{eq:higher_low_energy}
\begin{align}
    &\beta_0^n = r\mrm{H}^2/r\mrm{S}^2\,,\\
    &\beta_1^n = \dfrac{4r\mrm{H}^4}{r\mrm{S}^4}\left[\dfrac{\Gamma(1/(n+1))\Gamma(2/(n+1))}{(n+1)\Gamma(3/(n+1))}\right]^2\,,\\
    &\beta_2^n = \dfrac{16 r\mrm{H}^6}{r\mrm{S}^6}\left[\dfrac{\Gamma(1/(n+1)) \Gamma(4/(n+1))}{(n+1)\Gamma(5/(n+1))}\right]^2\,,\\
    &\beta_{1/2}^n = 2^{4-4/(n+1)}r\mrm{H}^2/r\mrm{S}^2\,.
\end{align}
\end{subequations}
We note that the low energy asymptotic limit for spin 2 is not given explicitly in \cite{Kanti2003_spin1_12}. However, their equations~(37) and (38) are valid for spin 2, as shows a careful follow-up of all the steps from the Teukolsky master equation. We also point out that in \cite{Creek2006} only the analytical results for spin 2 emission \textit{in the bulk} are given, while we focus here on \textit{brane} emission of massless gravitons. The results are shown in Fig.~\ref{fig:higher_asymptotics} (right panel). The comparison of the theoretical asymptotic limits and our numerically computed spectra once again shows very good agreement.

\subsection{Polymerized black holes}

\begin{figure}[h]
    \centering
    \includegraphics{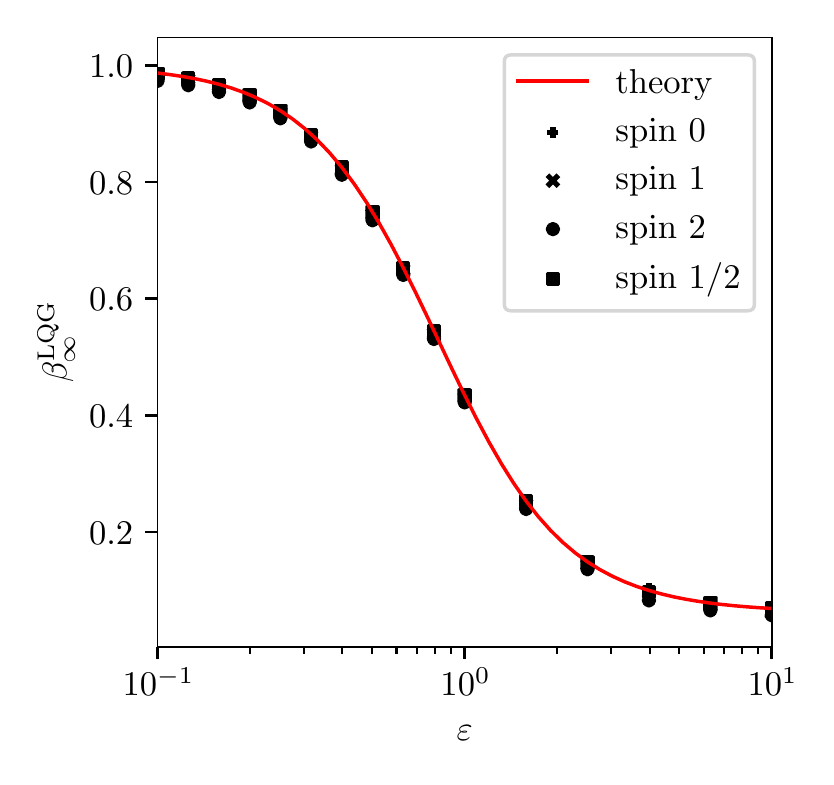}
    \includegraphics{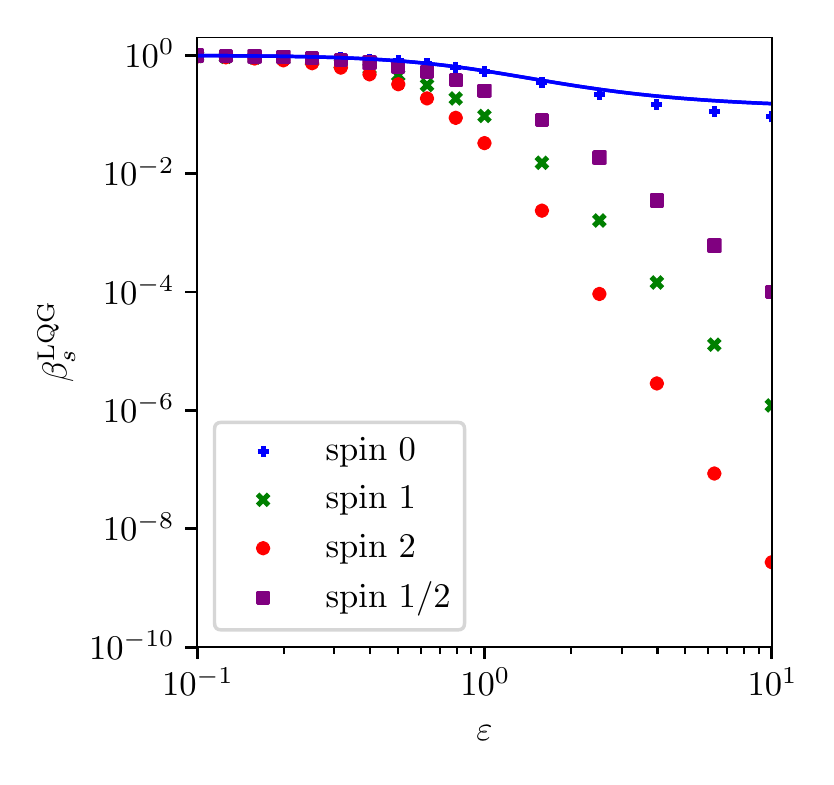}\vspace{-0.5cm}
    \caption{\textbf{Left:} Comparison between our high energy asymptotic limits for the cross-section of high energy fields with polymerized BHs (black markers), in the case $a_0 = 0$. The approximate formula \eqref{eq:LQG_high_energy} is shown as a red solid line. \textbf{Right:} Comparison between the low energy asymptotic cross-section of Eqs.~\eqref{eq:LQG_low_energy_spin0} for polymerized BHs (blue line) and our computed results for spin 0 (blue crosses), as well as the computed low energy cross-sections for the other spins (green, red and purple markers) for which literature is not available.}
    \label{fig:LQG_asymptotics_a0}
\end{figure}

The full HR spectra for massless fields of spins 0, 1, 2 and $1/2$ in the case of a polymerized BH described by the metric \eqref{eq:metric_LQG} are given in the Appendix~\ref{app:results} (see Figs.~\ref{fig:HR_LQG_a0} and \ref{fig:highe} for low and high values of $\varepsilon$ respectively). Most of these results are new, as we now explain.

In the literature, the spin 0 case has historically been treated first, theoretically in \cite{Alesci2011,Hossenfelder2012,Alesci2012} and numerically in \cite{Anacleto2020}. Then, spins 0 and $1/2$ have been studied in \cite{Moulin2019}, where in the spin $1/2$ case the Teukolsky equation is solved numerically without the intermediate step of deriving the short-ranged potential. Concerning the massless spin 2 perturbations (and other spins as well), they were only studied in the case of quasi-normal modes (QNMs) \cite{Barrau2019,Bouhmadi2020}. These studies invoke the same kind of potentials as in Eqs.~\eqref{eq:potentials}, but solve the Schr\"odinger-like wave equation with different boundary conditions to find the quasi-normal frequencies. It was not possible to compare our full spectra for low values of $\varepsilon$ for spins $0$ and $1/2$ to those of \cite{Moulin2019} due to unspecified normalization factors in their figures 2, 3 and 4. Moreover, is seems that we find results differing from theirs for massless fields of spin $1/2$: they predict a distortion of the spectra at high energies that we do not observe in the right panel of Fig.~\ref{fig:HR_LQG_a0} corresponding to the fermionic field. The height of the first peak in the spin $1/2$ cross-section seems to follow a different tendency in our results compared to theirs when $\varepsilon$ increases. However we find, as they do, that increasing the parameter $\varepsilon$ leads to a decreasing HR rate. A linear scale has been used in the left panels of Fig.~\ref{fig:HR_LQG_a0} to make this statement more obvious, while a logarithmic scale was used for high values of $\varepsilon$ in Fig.~\ref{fig:highe} as the HR is much more decreased. We also find that the parameter $a_0$ plays no particular role in the rates of emission, at least when varying between 0 and the often-used value $a_0 = \sqrt{3}\gamma/2 \simeq 0.11$ where $\gamma \equiv \ln(2)/\sqrt{3}\pi$ is the Barbero--Immirzi parameter, causing only a very slight decrease in the high energy tail due to the different temperatures. The differences are at the percentage level and not shown here. The effect of small values of $\varepsilon$ on the Hawking radiation rates is small altogether. 

The results are much more intriguing for high values of $\varepsilon$, and are reported on Fig.~\ref{fig:highe}. With values of $\varepsilon = \{1, 4, 10\}$ we obtain a reduction in the temperature of a factor $\sim \{1, 1.6, 3.2\}$ (respectively). The associated decrease in the emission rate is strongly spin-dependent, a feature that could be explained with an analysis of the precise dependency of the potentials in $\varepsilon$: the emission is more and more damped for high values of $\varepsilon$ as the particle spin increases. This is a major result which has important consequences on the Hawking radiation constraints discussed in the next Section. A fundamental difference between Reissner--Nordstr\"om (Kerr) and polymerized BHs is that the charge (angular momentum) is radiated away during the BH evaporation. Thus, the behaviour of the BH is expected to follow a Schwarzschild trajectory once these parameters are back to small values. The polymerization factor, on the other hand, is a constant inherited from the quantum nature of gravity. Thus, the specific behaviour associated with a high value of $\varepsilon$ (namely a decrease of the BH temperature compared to the Schwarzschild case, and smaller emission rates) should last during the whole lifetime of the BH. These aspects are discussed further in the next Section.

The high energy limit of the cross-section for polymerized BHs has not appeared in the literature. Using conditions~\eqref{eq:conditions}, we obtain the value for the radius of the photon sphere in the limit $a_0 \rightarrow 0$. This is
\begin{equation}
    r^{\rm LQG}\mrm{c}=\dfrac{r_+}{6}\left(3P^2-4P+3 + \dfrac{9 + P \left(6 + P (10 + 6 P + 9 P^2)\right)}{z}+z\right),
\end{equation}
where $z$ is
\begin{align}
    z \equiv \Bigg[ &27 + 27 P - 63 P^2 - 190 P^3 - 63 P^4 + 27 P^5 + 27 P^6 \nonumber\\
    &+  3 P (1 + P)\sqrt{-675 + 
   3 P (-238 + P (49 + P (636 + P (49 - P (238 + 225 P)))))}\Bigg]^{1/3}.
\end{align}
We therefore find that the high energy limit of the cross-section is given by
\begin{equation}
    \beta_\infty^{\rm LQG} = \left(\dfrac{r^{\rm LQG}\mrm{c}}{r\mrm{S}}\right)^2\dfrac{1}{G\big(r\mrm{c}^{\rm LQG}\big)}\,.\label{eq:LQG_high_energy}
\end{equation}
The comparison of this formula with our numerical results is shown in Fig.~\ref{fig:LQG_asymptotics_a0} (left panel), with excellent agreement even for high values of $\varepsilon$. In the case where $a_0\ne 0$, and in particular when taking the fiducial value $a_0 = \sqrt{3}\gamma/2 \simeq 0.11$ where $\gamma \equiv \ln(2)/\sqrt{3}\pi$ is the Barbero--Immirzi parameter \cite{Moulin2019}, we cannot use the approximate formula~\eqref{eq:LQG_high_energy}. However, we have performed a numerical estimation of the photon sphere radius $r\mrm{c}$ and compared the resulting high energy cross-section to our computed results, showing great agreement. The effect of taking $a_0\ne 0$ is small anyways. It was proven by \cite{Anacleto2020} that at low energy the scalar wave has a cross-section
\begin{equation}
    \beta_0^{\rm LQG} =\f{4m^2(1+P^2)}{r\mrm{S}^2}\left( 1 + \dfrac{a_0^2}{16m^4}\right)= \dfrac{r_+(r_+ + r_-)}{r\mrm{S}^2}\left( 1 + \dfrac{a_0^2}{r_+^4} \right).\label{eq:LQG_low_energy_spin0}
\end{equation}
The agreement of this analytical result with our computed results is shown in Fig.~\ref{fig:LQG_asymptotics_a0} (right panel). For the spin 0 field, the low $\varepsilon$ regime is correctly reproduced, but a discrepancy between the analytical limit and the numerical calculation is found at high values of $\varepsilon$. This difference is of order unity, and concerns only the very low energy asymptotic behaviour. It is not clear whether the derivation of the formula \eqref{eq:LQG_low_energy_spin0} is valid at high values of $\varepsilon$. There is, to our knowledge, no literature giving the asymptotic limits at low energy for the other spins. The very good agreement between these theoretical limits and our numerically computed results are convincing us that our potentials for this non-$tr$-symmetric example are efficient. We have checked that the limit $\varepsilon,a_0\rightarrow 0$ gives the Schwarzschild result for all spins. We point out that the high-$\varepsilon$ damping of the HR rate is very clear for spins $s > 0$, while the spin 0 emission rate is reduced by much.

\subsection{Constraints from polymerized black holes}
\label{sec:constraint}

In this Section, we obtain for the first time Hawking radiation constraints on polymerized primordial black holes using the results previously derived in this paper. There are two parameters at stake, $a_0$ and $\varepsilon$. First, $a_0$ is expected to be negligible for BHs with a radius $r_+^2 \gg a_0$, and to play a role only at the end of the BH evaporation, when its radius reaches values close to the Planck length, out of the Hawking radiation constraint range. The parameter $\varepsilon$, on the other hand, has an effect which is proportional to its value, with small values of $\varepsilon$ leading to very little changes in the Hawking radiation emission, while larger values may have a dramatic impact (see previous section). There are two major outcomes expected when considering the evaporation constraints on polymerized PBHs:
\begin{itemize}
    \item a decrease of the Hawking temperature and emission rates at high $\varepsilon$, which results in a longer lifetime, shifting the (time-dependent) constraints towards smaller PBH masses;
    \item this decrease will also lead to weaker (instantaneous) constraints.
\end{itemize}
Thus, the most striking result from this Section is that we expect the window for light PBHs to represent all DM to be reopened in the case of high values of $\varepsilon$, down to smaller PBH masses than in the Schwarzschild case.

\begin{figure}
    \centering
    \includegraphics{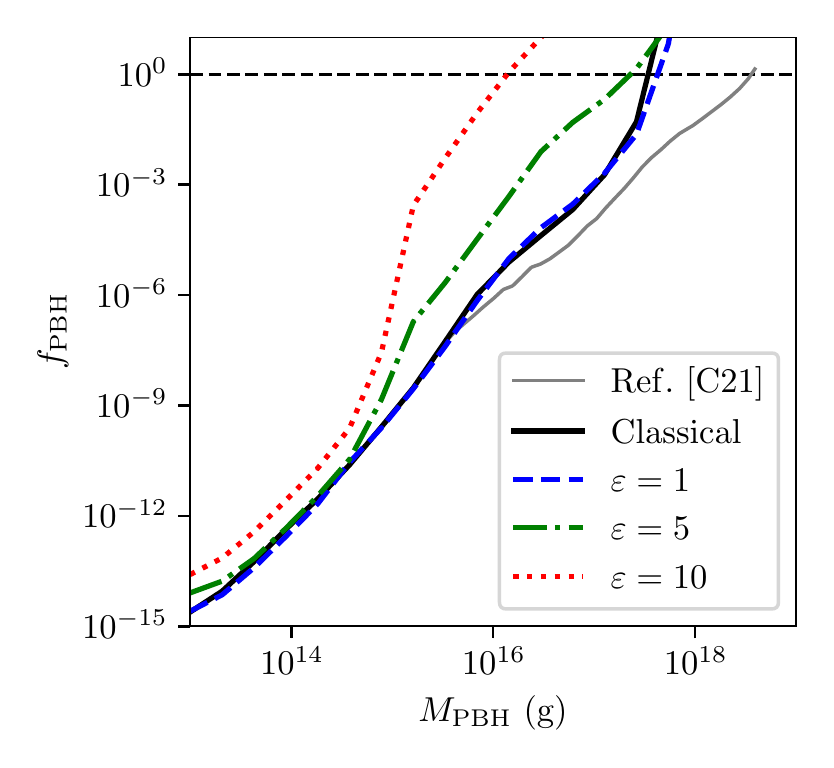}\vspace{-0.5cm}
    \caption{Constraints on the PBH fraction in DM from the measurement of MeV-GeV photons in the galactic center by AMEGO. We show the constraints derived for a classical BH in this work (solid black), to be compared to the same limit from \cite{Coogan2021} (solid grey, denoted as [C21]). Then we show the constraints computed for increasing values of $\varepsilon = \{1, 5, 10\}$ (dashed blue, dot dashed green and dotted red, from right to left). The horizontal dashed line denotes the limit $f\mrm{PBH} = 1$.}
    \label{fig:constraint}
\end{figure}

In order to illustrate this proposal, we have chosen to compute the prospective evaporation constraints from MeV to GeV photons as will be measured by AMEGO, whose expected sensitivity can be found in \cite{Amego2019}. There are two reasons for this choice: $i$) this limit lies among the most stringent ones in the disputed mass range where PBHs may represent all DM (see \cite{Coogan2021}), $ii$) some of the authors of this paper are also authors of the public code \texttt{BlackHawk}, which has been updated to compute precisely the secondary low energy photon spectra and obtain robust constraints in the considered mass range \cite{Coogan2021}, as well as to compute the primary emission rates for photons in the polymerized metric. We compute only the constraint from the AMEGO instrument for PBHs evaporating in the galactic center, as it is the most stringent one. The effects which we will describe below can also be applied to all the other evaporation constraints. We follow exactly the setup chosen by \cite{Coogan2021}: a NFW distribution of DM in the Milky Way, and observation in some small window of angular width $5^\circ$, which gives $\Delta\Omega = 2.39\times10^{-2}\,$sr. The expected emission is thus
\begin{equation}
    \dfrac{\d \Phi}{\d E} = \frac{1}{4\pi} \dfrac{f\mrm{PBH}}{M\mrm{PBH}} \dfrac{\d^2 N}{\d E\,\d t}  \int\mrm{LOS}  \rho\mrm{DM}\,\d l\,,
\end{equation}
where the integral over the line of sight can be written as
\begin{equation}
    J \equiv \dfrac{1}{\Delta \Omega} \int_{\Delta \Omega} \d\Omega \int_{\rm LOS} \rho\mrm{DM}\,\d l = 1.597\times10^{26}\,\text{MeV}\cdot\text{cm}^{-2}\cdot\text{sr}^{-1}\,.
\end{equation}
The numerical value comes from Refs.~\cite{Coogan2021,Salas2019}. The results are shown in Fig.~\ref{fig:constraint} where we plot the constraints for $\varepsilon = \{1,5,10\}$ as well as the fiducial constraint for the classical case and the constraint from \cite{Coogan2021} for comparison. The constraints are obtained by maximizing $f\mrm{PBH}$ while keeping $E^2 \d\Phi/\d E$ below the AMEGO sensitivity (figure 5 of \cite{Amego2019}).

We observe that the constraints we derive for the classical Schwarzschild case have differences with the results of \cite{Coogan2021}. We were not able to explain their origin, since we have carefully reproduced their low energy photon spectra and use the same source for the AMEGO sensitivity. As expected, the constraints for the classical Schwarzschild BH and the polymerized BH with $\varepsilon = 1$ (small polymerization factor) are similar, as their Hawking radiation rates are very close (see Fig.~\ref{fig:HR_LQG_a0}). Then, as we increase $\varepsilon$ to 5 and then 10, we observe that the constraints get weaker in the high mass range $M\mrm{PBH} \gtrsim 10^{15}\,$g, allowing the DM fraction $f\mrm{PBH}$ of PBHs to be 1 for $M\mrm{PBH}\gtrsim 10^{17}\,$g (or $10^{16}\,$g for $\varepsilon=10$). This is due to the fact that the main contribution to the photon spectra for these PBHs comes from the directly emitted primary photons, whose emission rate is strongly suppressed when $\varepsilon$ increases (see Fig.~\ref{fig:highe}). However, in the lower mass range $M\mrm{PBH} \lesssim 10^{15}\,$g, the constraints remain of the same order of magnitude. In this energy range, the constraints come from the secondary photons generated by neutral pion decay. As pions are spin 0 particles, their emission rate decreases slowly as $\varepsilon$ increases (see Fig.~\ref{fig:highe}); the effect of the polymerization factor becomes sizeable only for extreme values. In this constraint plot, we have extended the mass range to masses $M\mrm{PBH} = 10^{13}\,$g, which is 2 orders of magnitude below the usual evaporation limit $M\mrm{PBH}\lesssim 10^{15}\,$g set by the lifetime of the PBHs, because we expect that the decreased emission rates will result in an increased PBH lifetime, thus allowing smaller PBHs to contribute to DM today. This effect will be quantitatively studied in future work, as well as the modifications of the other set of constraints (\textit{e.g.}~electrons \cite{DeRocco2019,Laha2019} and time-stacked extra-galactic background \cite{Arbey2020_EGRB}).

One last aspect, which we have not quantitatively explored, is the rate of final bursts of PBHs and their observation by gamma ray instruments (see the recent paper \cite{Doro2021}). Since  in the high $\varepsilon$ limit PBHs evaporating today would have a lower initial mass than in the Schwarzschild case, their number abundance should be larger if they represent some fixed fraction of DM. Thus, the rate of nearby final bursts would be higher, leading to more stringent constraints. However, the non-trivial modification of the final light curves (energies, duration) makes it difficult to predict the sensitivity of the gamma ray instruments to these polymerized PBHs burst.

\section*{Conclusion}
\label{sec:conclusion}

In this paper we have used the short-ranged potentials derived in the companion paper \cite{Arbey:2021jif} to compute numerically the Hawking radiation signals from three spherically-symmetric and static BH solutions: charged BHs, higher-dimensional BHs and polymerized BHs. We have checked the robustness of our results by comparing the HR at low and high energy with analytical formulas, some of which have been derived here for the first time. Focusing on the case of polymerized BHs, whose peculiar metric form is the heart of our analytical study, we conclude that the HR signals are not much different from the Schwarzschild case in the limit $\varepsilon \ll 1$ of small polymerization parameter. We have shown however that if the polymerization parameter takes high values $\varepsilon \gtrsim 1$, then all the evaporation constraints would need to be re-evaluated with two major effects: shifting the constraints towards smaller PBH masses, and relaxing the constraints on the fraction of DM which light PBHs can represent. The modification of the constraints depends in a non-trivial way on the spin of the primary particles involved in the secondary observed spectrum; the effect of the quantum deformation $\varepsilon$ is very spin-dependent. The main consequence of this result is that the mass range usually excluded by (future) evaporation limits $10^{16}\,{\rm g} < M <10^{18}\,{\rm g}$ for all DM in the form of PBHs is reopened. This is a striking result in PBH DM studies, and the first constraint ever set on the fraction of polymerized primordial black holes as dark matter using Hawking radiation signals. As a final remark, we have opened a new window on the study of HR from other regular BH metrics, which is an exciting prospect.

%%%%%%%%%%%%%%%%%%%%%%%%%%%%%%%%%%%%%%%%%%%%%%%%%%%%%%%%%%%%%%%%%%%%%%%%

\newpage

\appendix

\section{Tortoise coordinates}
\label{sec:tortoise}

In order to solve the Schr\"odinger wave equation \eqref{eq:schrodinger} with the potentials \eqref{eq:potentials}, we need to translate from the radial coordinate to the tortoise coordinate. Here we give the results of the analytical integration of Eq.~\eqref{eq:tortoise} for the three metrics considered in Section~\ref{sec:metrics}.

\paragraph{\textbf{Charged black holes.}}
\label{sec:tortoise_charged}

For the metric \eqref{eq:metric_charged}, integration of Eq.~\eqref{eq:tortoise} gives
\begin{equation}
    r^*(r) = r + \dfrac{r_+^2}{r_+ - r_-}\ln\left( \dfrac{r}{r_+} - 1 \right) - \dfrac{r_-^2}{r_+ - r_-}\ln\left( \dfrac{r}{r_-} - 1 \right).
\end{equation}
with $r_\pm$ as in Eq.~\eqref{eq:rpm_charged}.

\paragraph{\textbf{Higher-dimensional black holes.}}
\label{sec:tortoise_higher}

For the metric \eqref{eq:metric_higher}, integration of Eq.~\eqref{eq:tortoise} gives for $n = 1$
\begin{equation}
    x_1^*(x) = x + \dfrac{1}{2}\ln\left( \dfrac{x - 1}{x + 1} \right),
\end{equation}
for $n = 2$
\begin{equation}
    x_2^*(x) = x + \dfrac{1}{3}\ln(x - 1) + \dfrac{1}{\sqrt{3}}\arctan\left( \dfrac{\sqrt{3}}{1 + 2x} \right) - \dfrac{1}{6}\ln\left( x^2 + x + 1 \right),
\end{equation}
for $n = 3$
\begin{equation}
    x_3^*(x) = x + \dfrac{1}{4}\ln\left( \dfrac{x-1}{x+1} \right) + \dfrac{1}{2}\arctan\dfrac{1}{x}\,,
\end{equation}
for $n = 4$
\begin{align}
    x_4^*(x) = x &+ \dfrac{1}{5}\ln(x-1) +  \dfrac{\sqrt{\sqrt{5}}}{5}\left[ \sqrt{\varphi_+}\arctan\left( \dfrac{\sqrt{\sqrt{5}}\sqrt{\varphi_+}}{2x + \varphi_-} \right) + \sqrt{\varphi_-}\arctan\left( \dfrac{\sqrt{\sqrt{5}}\sqrt{-\varphi_-}}{2x + \varphi_+} \right) \right] \nonumber\\
	&-\dfrac{1}{10}\left[ \varphi_-\ln(x^2 + \varphi_- x + 1) + \varphi_+\ln(x^2 + \varphi_+ x + 1)\right],
\end{align}
where $\varphi_\pm \equiv (1\pm\sqrt{5})/2$, for $n = 5$
\begin{equation}
    x_5^*(x) = x + \dfrac{1}{6}\ln\left( \dfrac{x - 1}{x + 1} \right) + \dfrac{1}{2\sqrt{3}}\left[ \arctan\left(  \dfrac{\sqrt{3}}{2x + 1}\right) + \arctan\left( \dfrac{\sqrt{3}}{2x - 1} \right) \right] +\dfrac{1}{12}\ln\left( \dfrac{x^2 - x + 1}{x^2 + x + 1} \right),
\end{equation}
and for $n = 6$
\begin{align}
    &x_6^*(x) = x + \dfrac{1}{7}\ln\left( x - 1 \right)\nonumber \\
    &+ \dfrac{2}{7}\Bigg\lbrack \cos\left( \pi/14 \right)\arctan\left(  \dfrac{\cos(\pi/14)}{x + \sin(\pi/14)}\right) + \cos\left( 3\pi/14 \right)\arctan\left( \dfrac{\cos(3\pi/14)}{x - \sin(3\pi/14)} \right) 
    + \sin\left( \pi/7 \right)\arctan\left( \dfrac{\sin(\pi/7)}{x + \cos(\pi/7)} \right) \Bigg\rbrack \nonumber\\
    &+ \dfrac{1}{7}\Bigg\lbrack \sin\left( 3\pi/14 \right)\ln(x^2 - 2\sin(3\pi/14)x + 1) - \sin\left( \pi/14 \right)\ln(x^2 + 2\sin(\pi/14)x + 1) - \cos\left( \pi/7 \right)\ln(x^2 + 2\cos(\pi/7)x +1) \Bigg\rbrack\,.
\end{align}
In these relations, we have defined $x^* \equiv r^*/r\mrm{H}$ and $x \equiv r/r\mrm{H}$.

\paragraph{\textbf{Polymerized black holes.}}
\label{sec:tortoise_LQG}

For the metric \eqref{eq:metric_LQG}, integration of Eq.~\eqref{eq:tortoise} gives
\begin{equation}
    r^*(r) = r - \dfrac{a_0^2}{r_+r_- r} + \dfrac{a_0^2(r_++r_-)}{r_+^2 r_-^2}\ln\left( \dfrac{r}{r_++r_-} \right) + \dfrac{a_0^2+r_+^4}{r_+^2(r_+-r_-)}\ln\left(\dfrac{r}{r_+} - 1\right) + \dfrac{a_0^2+r_-^4}{r_-^2(r_- - r_+)}\ln\left(\dfrac{r}{r_-} - 1\right)\,.
\end{equation}

\section{Detailed results for Hawking radiation}
\label{app:results}

The various plots mentioned in the core of the article are gathered below. They feature the different spins $(0,1,2,1/2)$ from top to bottom.

%\subsection{Charged BHs}

\begin{figure}[!]
    \centering
    \includegraphics{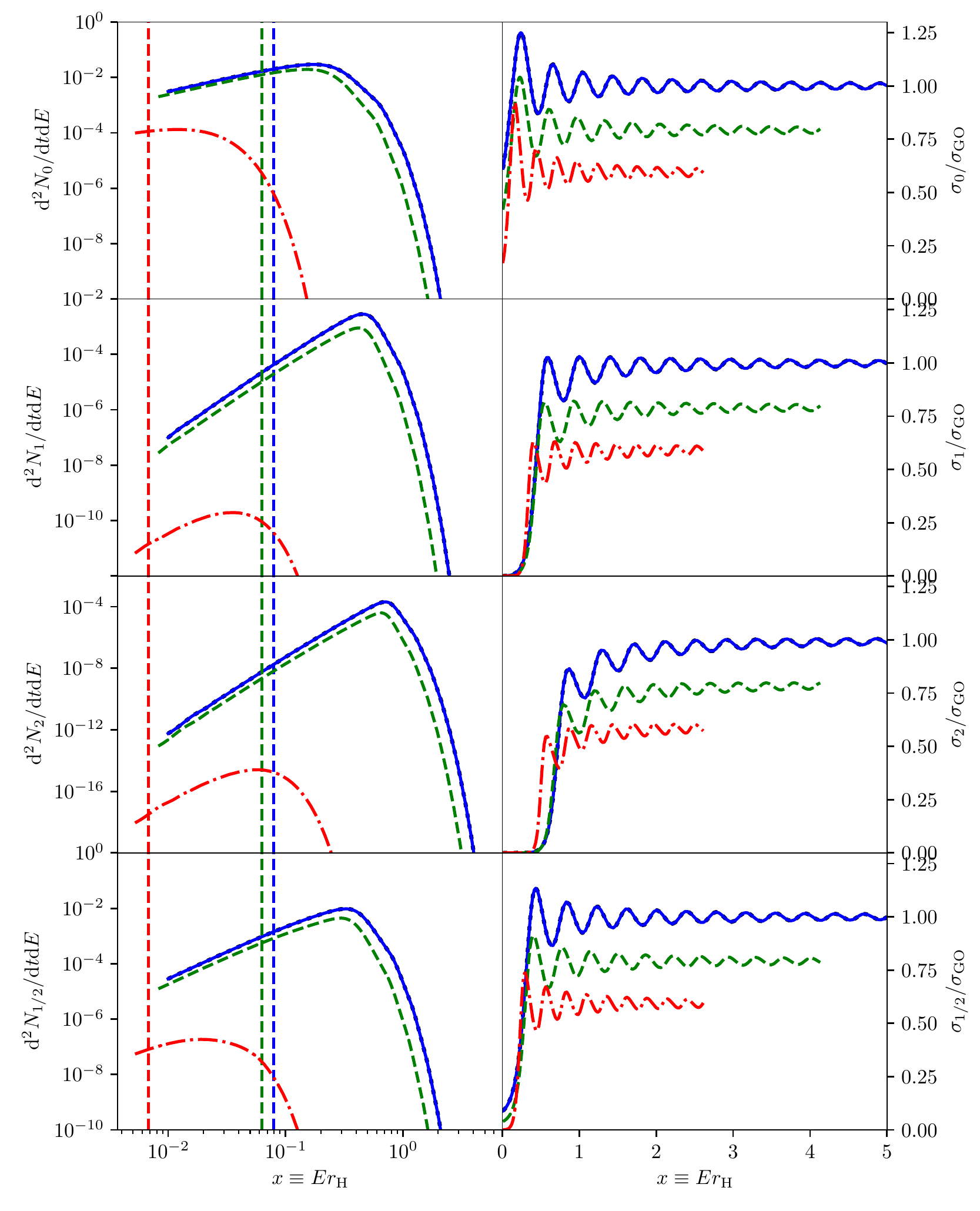}
    \caption{Hawking radiation of massless particles of spin $(0,1,2,1/2)$ (top to bottom) from charged BHs with $Q = \{0.010,\, 0.758,\, 0.999\}M$ (solid blue, dashed green and dot-dashed red respectively). The Schwarzschild BH with $Q = 0$ is in dotted black. The $Q = 0.010$ curves are indistinguishable from the Schwarzschild ones. The vertical lines on the left panels represent the temperature of the BH. Be careful of the different $x$ and $y$ axes values.}
    \label{fig:HR_charged}
\end{figure}

%\subsection{Higher-dimensional BHs}

\begin{figure}[!]
    \centering
    \includegraphics{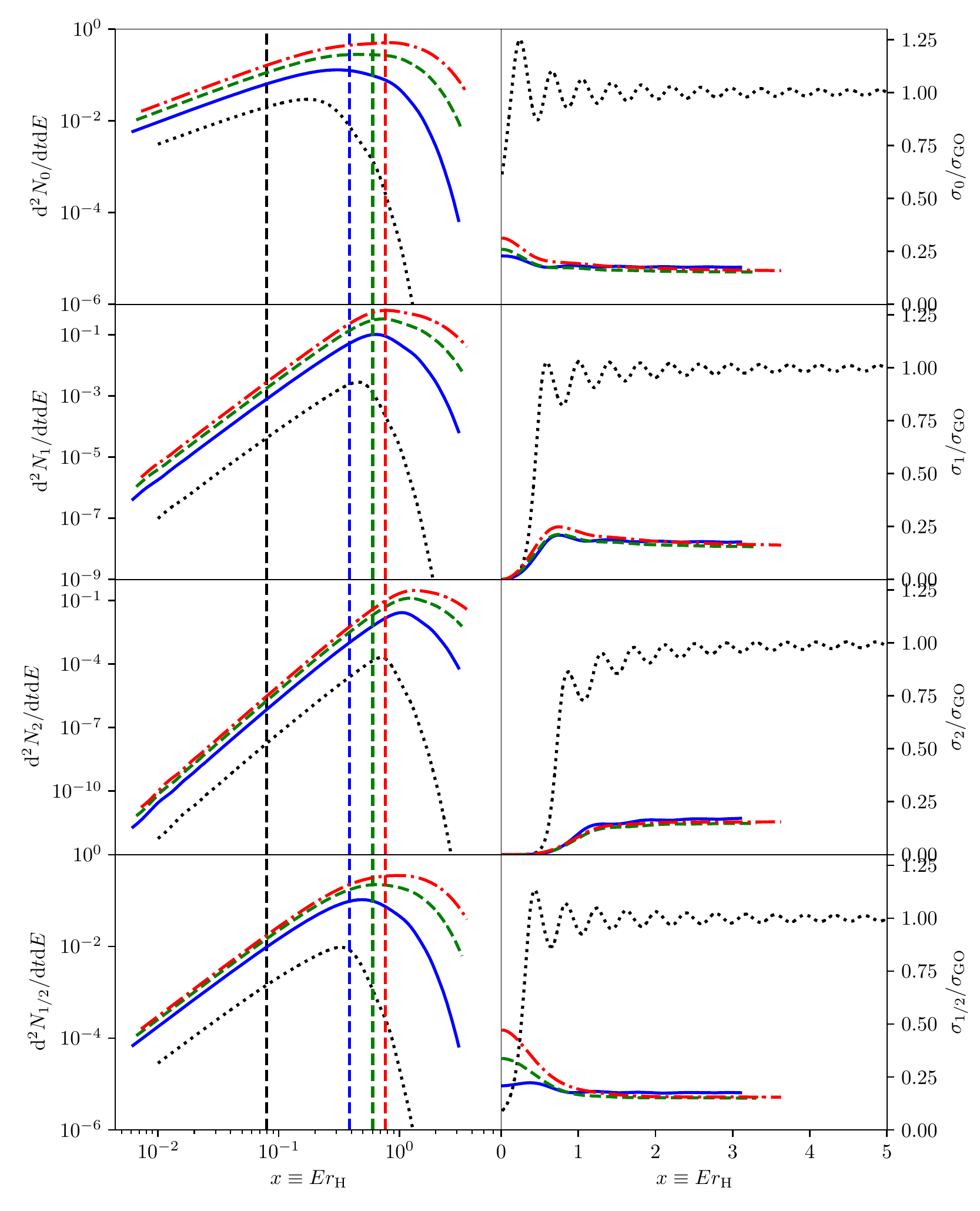}
    \caption{Hawking radiation of massless particles of spin $(0,1,2,1/2)$ (top to bottom) from higher-dimensional BHs with $n = \{2,\, 4,\, 6\}$ (solid blue, dashed green and dot-dashed red respectively) and $M^* = 1$. The Schwarzschild BH with $n = 0$ is in dotted black. The vertical lines on the left panels represent the temperature of the BH. Be careful of the different $x$ and $y$ axes values.}
    \label{fig:HR_higher}
\end{figure}

%\subsection{Polymerized BHs}

%\begin{figure}
%    \centering
%    \includegraphics{spin_all_LQG.pdf}
%    \caption{Comparison of Hawking radiation for all massless particles spin from polymerized BHs with $\varepsilon = \{10^{-1},\, 10^{-0.6},\, 10^{-0.1}\}$ (solid blue, dashed green and dot-dashed red respectively) and $a_0 = 10^{-10}$ to a Schwarzschild BH (dotted black). The vertical lines on the left panels represent the temperature of the BH, which are indistinguishable within our choice of parameters. Be careful of the different $x$ and $y$ axes values and scales.}
%    \label{fig:HR_LQG}
%\end{figure}

\begin{figure}[!]
    \centering
    \includegraphics{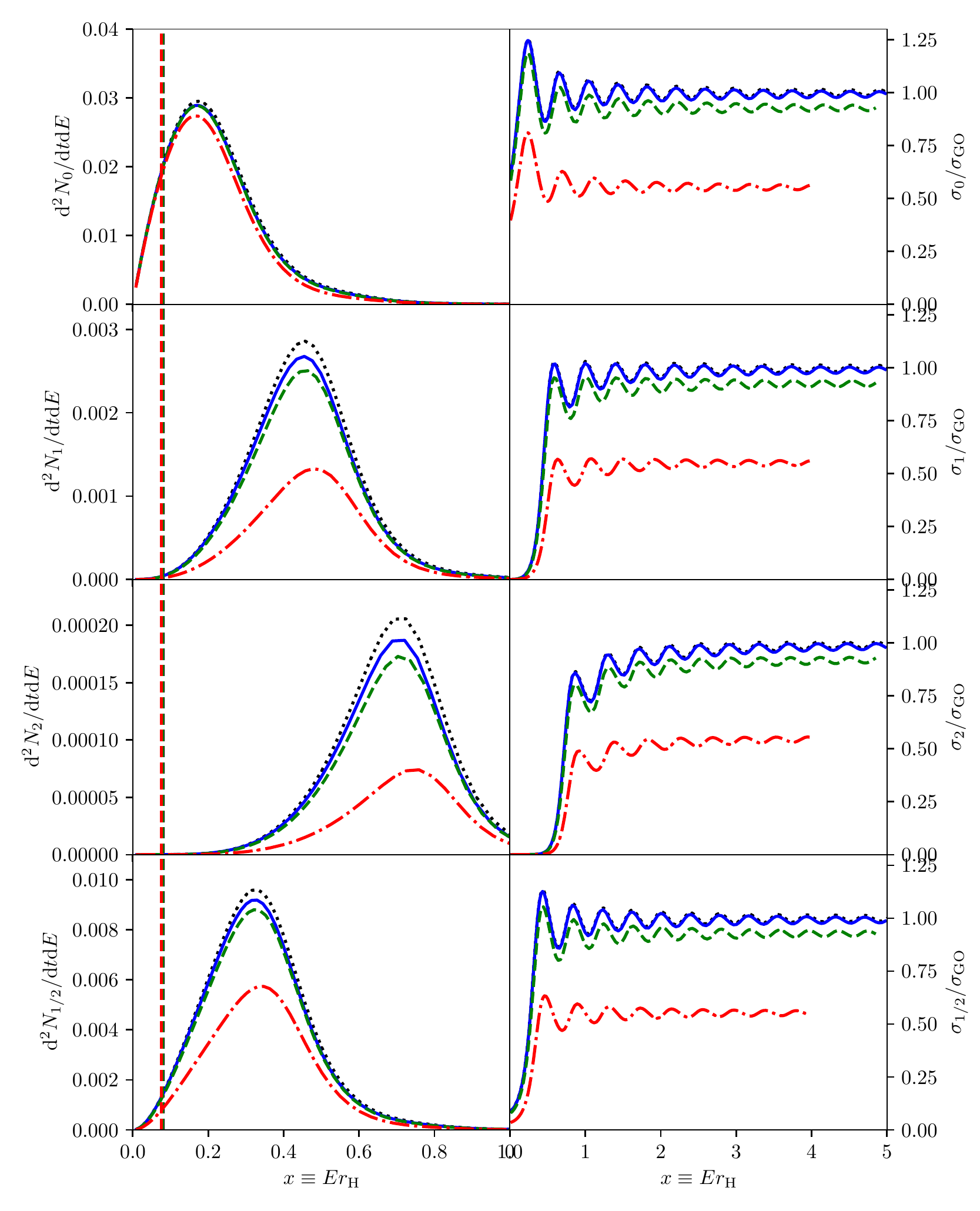}
    \caption{Hawking radiation of massless particles of spin $(0,1,2,1/2)$ (top to bottom) from polymerized BHs with $\varepsilon = \{10^{-1},\, 10^{-0.6},\, 10^{-0.1}\}$ (solid blue, dashed green and dot-dashed red respectively) and $a_0 = \sqrt{3}\gamma/2$ where $\gamma \equiv \ln(2)/\sqrt{3}\pi$ is the Barbero--Immirzi parameter \cite{Moulin2019}. The Schwarzschild BH is in dotted black. The vertical lines on the left panels represent the temperature of the BH, which are indistinguishable within our choice of parameters. Be careful of the different $x$ and $y$ axes values and scales.}
    \label{fig:HR_LQG_a0}
\end{figure}

\begin{figure}[!]
    \centering
    \includegraphics{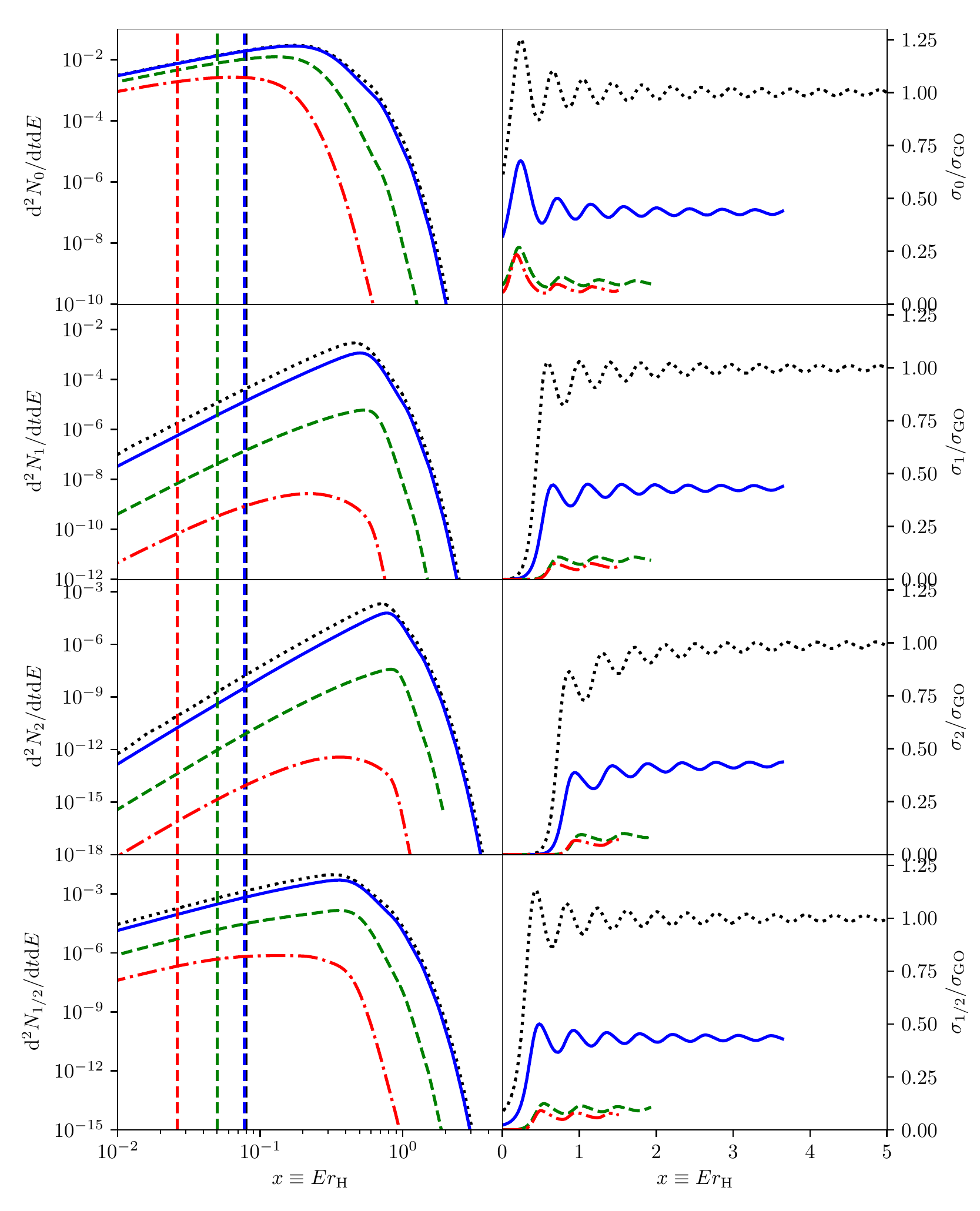}
    \caption{Hawking radiation of massless particles of spin $(0,1,2,1/2)$ (top to bottom) from polymerized BHs with $\varepsilon = \{1,\, 4,\, 10\}$ (solid blue, dashed green and dot-dashed red respectively) and $a_0 = 0$. The Schwarzschild BH is in dotted black. The vertical lines on the left panels represent the temperature of the BH. Be careful of the different $x$ and $y$ axes scales.}
    \label{fig:highe}
\end{figure}

\newpage

\bibliography{biblio}

\end{document}